\newcommand{\beq}{\begin{equation}}
\newcommand{\eeq}{\end{equation}}
\documentstyle[eqsecnum,preprint,aps,epsf]{revtex}
\begin{document}
\draft
\preprint{IC/97/54}
\title{
Analysis of the Interplay of Quantum Phases and
Nonlinearity Applied to Dimers with Anharmonic
Interactions}

\author{S. Raghavan}
\address{Condensed Matter Section, 
International Centre for Theoretical Physics\\ 
I-34100 Trieste, Italy}

\maketitle
\begin{abstract}
We extend our analysis of the effects of the interplay of
quantum phases and nonlinearity to address saturation effects in
small quantum systems.
 We find that initial phases dramatically
control the dependence of self-trapping on initial asymmetry of
quasiparticle population and can compete or act with nonlinearity as
well as saturation effects.
We find that there is a  minimum
finite saturation
value in order to obtain self-trapping that crucially
depends on the initial
quasiparticle phases and present a detailed phase-diagram in terms
of the control parameters of the system: nonlinearity and saturation.
\end{abstract}

\renewcommand{\theequation}{\arabic{section}.\arabic{equation}}
\setcounter{equation}{0}
\section{Introduction}
We analyze here some features of a quasiparticle interacting
with lattice vibrations that are strongly anharmonic
in nature. An example of such an oscillator is to be found
in liquid crystals where the vibration is rotational \cite{chandra}.
Work done on
these lines by Kenkre and collaborators on {\em rotational polarons}
has shown the existence of fascinating phenomena like
saturation of selftrapping on {\em increasing} nonlinearity
beyond a characteristic value \cite{samos,kwh,wk}. Further work on
a different kind of oscillator, one which is harmonic for small
displacements from equilibrium but subjected to a restoring force, which
is logarithmic for large displacements also showed similar
behaviour \cite{occassnot}. In a study of the more commonly encountered
harmonic systems, it was shown
\cite{interplay}
that the initial quantum phases can profoundly control the process of
self-trapping in nonlinear quantum dimers where the interacting vibrations
were assumed to be harmonic (see also \cite{tk}).
The purpose of this note is to
report some interesting behaviour that occurs when the initial quantum
phases can interact with the nonlinearity in the system as well as
the saturation effects of anharmonic vibrations. We point out here that
throughout this paper, we shall assume the
vibrations to be classical.
It is important to note
that, although the regime
of validity of the DNLSE as a consequence of
microscopic dynamics is limited
\cite{vit,sbkr,krbs}, the problem of the interplay between
qunatum phases and nonlinearity studied in this paper is of
general interest in quantum nonlinear equations of
evolution.
The results of the
present paper have to be interpreted in this context.We emphasize that,
throughout this paper, as in earlier
work \cite{samos,kc,scatter,ktc,tk,kt}, by the
phrase `quantum phases', we refer only to the quantum mechanical phases
associated with the quasiparticle. Clearly, after the semiclassical
approximation has been made, there is no meaning to the association of
a quantum phase with the vibrational variables.
\par
We begin here with the following coupled equations \cite{samos,kwh,wk} for the
generalized displacement $x_m$ of the oscillator at site $m$, and for
the quasiparticle amplitude $c_m$ at site $m$,
\begin{eqnarray}
i\hbar\frac{dc_m}{dt} = \sum_n V_{mn} c_n + E(x_m) c_m, \\
\frac{d^2x_m}{dt^2} + \omega^2 f(x_m) + RE'(x_m)|c_m|^2 = 0,
\end{eqnarray}
where $V_{mn}$ is the intersite matrix element between sites $m$ and $n$,
$R$ is a proportionality constant related to the geometry of the system,
$E(x_m)$ is a generally nonlinear function of the coordinate $x_m$
describing the quasiparticle-vibration interaction, and $\omega^2 f(x_m)$
is a general  nonlinear function giving the restoring force on the
vibrational coordinate.

\par
For the rotational polaron, as noted elsewhere \cite{samos,kwh,wk},
the restoring force and the interaction are described as
\begin{equation}
f(x) = \frac{\sin(\Lambda x)}{\Lambda}, \; E(x) = \frac{E_0}{\Lambda}
\sin(\Lambda x).
\label{eq:rotpol:fE}
\end{equation}
Following the analysis of Kenkre et al. \cite{kwh,wk}, one obtains
the amplitude equation
\begin{equation}
i\hbar\frac{dc_m}{dt} = \sum_n V_{mn} c_n -
\frac{\chi |c_m|^2}{\sqrt{1+(\chi/\Delta)^2|c_m|^4}} c_m.
\label{eq:gdnls:rotpol}
\end{equation}
in general, and for the two-site system (dimer) in particular,
\begin{equation}
\dot{p} = -2Vq \;,\;\dot{q}=2Vp + \frac{dg(p)}{dp}r\;,\;
\dot{r} = -\frac{dg(p)}{dp}q,
\label{rotpol:pqr=}
\end{equation}
where
\begin{equation}
g(p) = \Delta\left[\sqrt{(p+1)^2 + \left(\frac{2\Delta}{\chi}\right)^2} +
\sqrt{(p-1)^2 + \left(\frac{2\Delta}{\chi}\right)^2} -% \\ \nonumber
2\sqrt{1+\left(\frac{2\Delta}{\chi}\right)^2}\right].
\label{rotpol:gofp}
\end{equation}
with the definitions
\begin{equation}
p = |c_1|^2 - |c_2|^2 \; ; \;
q = i (c_{1}^{*} c_{2} - c_{1} c_{2}^{*}) \; ; \;
r = c_{1}^{*} c_{2} + c_{1} c_{2}^{*}.
\label{pqr}
\end{equation}
and where $p_0,r_0$ refer to the initial values of $p,r$ respectively.
Here the nonlinearity $\chi=E_0^2 R/\omega^2$ and the saturation energy
$\Delta=E_0/\Lambda$.

As has been shown in Refs.~\cite{samos,kwh,wk}, one obtains from
Eq.~(\ref{rotpol:gofp})
an equation of motion for a
fictitious conservative oscillator whose displacement
is $p$, and that moves in a potential given by
\begin{equation}
U(p) = 4V^2p^2 + g^2(p) - 2g(p)[g(p_0) - 2Vr_0],
\label{Uofp}
\end{equation}

\renewcommand{\theequation}{\arabic{section}.\arabic{equation}}
\setcounter{equation}{0}
\section{Rotational polaron}
We start this section with the elementary observation that
as long as $g(p)$ in Eq.~(\ref{Uofp}) is an even function of $p$,
the potential will also be one. Furthermore, if $g(0)=0$, $U(0)=0$.
This implies that the critical condition
 to determine the transition between
free evolution to  self-trapped behaviour can be found
 by examining the
condition $U(p_0) \leq 0$.
This condition combined with Eq.~(\ref{Uofp}) gives
\begin{equation}
|g(p_0)-2Vr_0| \geq 2V.
\label{free-self}
\end{equation}
It is therefore instructive to view the condition expressed
in Eq.~(\ref{free-self}) in terms of a phase-diagram with
the saturation ratio $\Delta/V$ and the critical nonlinearity ratio
$\chi_c/V$ as the parameters
for different values of $p_0,r_0$ marking regions of
free and self-trapped evolution.
We first examine the case for real initial conditions, i.e.,
$q(0) \equiv q_0 = 0$. In the next section, we shall generalize
to complex initial conditions.
\par
Figure 1 shows the phase diagram for the case when $r_0 < 0$.
It is to be interpreted in the following manner. For a given value of
the saturation ratio $\Delta/V$, as one increases the
value of nonlinearity ratio $\chi_c/V$, one traverses
regions of free-evolution. As soon as one encounters the boundary (for the
relevant value of $p_0$), one enters the region of self-trapped evolution.
This is the region where Eq.~(\ref{free-self}) is satisfied.
Note that upon increasing the value of nonlinearity beyond a certain value
one crosses this region and passes into free evolution again. This is
as expected and is in accordance with the observation in
Refs.~\cite{kwh,wk} that self-trapping can be destroyed by increasing
the nonlinearity. Note also that for the initial phase chosen so that
$r_0 < 0$, for a given value of saturation,
 it requires less nonlinearity
to enter the self-trapped phase if the amount of initial asymmetry
(measured by $p_0$) is less. This is in consonance with the observation
of Ref.~\cite{interplay}. It is interesting that this interplay between
choice of initial
phase and initial asymmetry remains qualitatively unaffected even in
the presence of saturation effects.
It is important, also, to note that
one gets out of this self-trapped phase for correspondingly lower values
of nonlinearity. It is interesting to see that as one moves to lower
values of $p_0$, the system tolerates lesser values of saturation to enter
the self-trapped phase, the most stringent limit being posed at the
value $p_0=1$.
 It is straightforward to estimate the
least saturation ratio required (and the corresponding value of
critical value of nonlinearity ratio):
 $\Delta_c/V = \sqrt{27/4}, \chi_c/V = \sqrt{27/2}$. It occurs
for $p_0 \rightarrow 0$ and is in agreement with what is obtained
in Ref.~\cite{wk}.  There are several features of interest in the
phase diagram. For fixed $p_0$, the phase boundary exhibits two
asymptotes. The first of these occurs for finite $\chi_c/V$ and
$\Delta/V \rightarrow \infty$. This is the standard DNLSE limit and
it corresponds to $\chi \ll \Delta$. The value of this asymptote
is given by $\chi_c/V = 4(1 - \sqrt{1-p_0^2})/p_0^2$ and is the
 result of Ref.~\cite{interplay}. The second asymptote is actually more
interesting since it marks the boundary marking the destruction of
self-trapping. The behaviour along this boundary is given by the
relationship
\begin{equation}
\frac{\chi_c}{V} = \left[\frac{\Delta}{V}\right]^{3/2}\alpha(p_0),
\label{rotpol:r0negasymp}
\end{equation}
where $\alpha(p_0) = p_0\sqrt{2}/[(1-p_0^2)(1-\sqrt{1-p_0^2})]^{1/2}$.
Equation (\ref{rotpol:r0negasymp}) shows
 that for small and intermediate values of
$p_0$, the dependence of $\chi_c/V$ on $\Delta/V$ is of power-law type
with an exponent of 1.5. The case of $p_0=1$ has to be dealt with
carefully, however, and an analysis shows that when $p_0 \rightarrow 1$,
the equation for the boundary becomes $\chi_c/V = (\Delta/V)^2$. The
exponent thus
changes from 1.5 to 2 and this explains the clear gap in
the upper phase boundary between small and large values of $p_0$.
\par
Let us now examine the case $r_0 > 0$. Figure 2 is the
$r_0 > 0$ counterpart of Fig.~1. One immediately notices the
profound difference between the two cases $r_0 < 0$ and $r_0 > 0$.
For instance,
one has the largest available area of parameter space for the self-trapped
phase when $p_0=1$ and this area progressively decreases when $p_0$ decreases
from this value. In contrast to the case where $r_0 < 0$, the phase
boundaries for different values of $p_0$ lie embedded within one another
in a non-overlapping fashion. It is instructive and straightforward to
calculate asymptotic values of the various features of the phase diagram
especially in the region of small $p_0$. In this limit, the turning point
of the phase boundary (marking the mininum amount of staturation allowed for
self-trapping) scales with the initial population difference as
$\Delta_c/V \sim 6\sqrt{3}/p_0^2$ and the corresponding value of
critical nonlinearity is $\chi_c/V = \sqrt{2} \Delta_c/V$. Again, in this
limit, there are two asymptotes for the phase-boundary. The first
occurs when one is in the standard DNLSE regime and
$\chi \ll \Delta$. Here,
one recovers the standard result that the critical value of nonlinearity
needed to trap, $\chi_c/V = 8/p_0^2$ and $\Delta/V \rightarrow \infty$.
The second asymptote marks the boundary where one gets out of the
self-trapped phase to the freely evolving phase. This is in the strongly
anharmonic phonon regime. The relationship between the parameters at this
boundary is markedly different from the asymptote discussed above and the
following equation for the boundary holds:
 $\chi_c/V \approx p_0 (\Delta/V)^{3/2}$. It is important to note that
this power-law dependence of $\chi_c/V$ on $\Delta/V$ is the similar to the
one found for the case $r_0 < 0$. However, the crucial point is that
when $r_0 < 0$,
the dependence of $\chi_c/V$ and $\Delta/V$ on $p_0$ is quite weak,
particularly for $p_0$ far from 1. On the other hand, for $r_0 > 0$, both
$\chi_c/V$ and $\Delta/V$ diverge as $1/p_0^2$. This explains why,
whereas for $r_0 < 0$,
on the upper phase boundary, the
 curves seem to merge for different values of $p_0$, the corresponding curves
are well-separated for $r_0 > 0$. There is thus a clear quantitative
difference between the two quantum phases $r_0<0$ and $r_0>0$.

\renewcommand{\theequation}{\arabic{section}.\arabic{equation}}
\setcounter{equation}{0}
\section{Complex initial amplitudes}
Thus far, we have only examined the case where the initial
conditions for the quasiparticle
are real, i.e., $q_0 = 0$. However, it is important to understand
the effect of arbitrary initial quantum phases. To this end, we write
the basic condition for examining the self-trapped-free phase boundary
(\ref{free-self}) as
\begin{equation}
|g(p_0)\pm 2V\sqrt{1-p_0^2-q_0^2}| \geq 2V.
\label{free-self:complex}
\end{equation}
Using the expression (\ref{rotpol:gofp}) in Eq.~(\ref{free-self:complex})
above, we plot in Fig.~3, the parameter space indexed by the
saturation ratio $\Delta/V$ and the nonlinearity $\chi_c/V$ separating
the self-trapped regions from the free region. For both Fig.~3(a) and
Fig.~3(b),
$r_0<0$. In Fig.~3(a), $p_0=0.1$ and in Fig.~3(b), $p_0=0.8$.
The primary observation is that relaxing the constraint of reality of
the initial quantum amplitudes 
results in a loss of parameter space available
for self-trapping. For small values of $p_0$ (illustrated by Fig.~3(a)),
the phase boundary is much more sensitive to changes in $q_0$. In fact,
for large and intermediate values of $q_0$ and for small values of $p_0$,
the diagram looks very much like the one obtained in Fig.~2 for $r_0 > 0$.
This can be understood very easily by analyzing the expression
(\ref{rotpol:gofp}) in conjunction with the equality condition of
Eq.~(\ref{free-self:complex}) that results in
\begin{equation}
\frac{\Delta}{V} = \frac{2(1-[1-p_0^2-q_0^2]^{1/2})}
{\sqrt{(p_0+1)^2 + \left(\frac{2\Delta}{\chi}\right)^2} +
\sqrt{(p_0-1)^2 + \left(\frac{2\Delta}{\chi}\right)^2} -
2\sqrt{1+\left(\frac{2\Delta}{\chi}\right)^2}}.
\label{eq:cmplxrtplbndry}
\end{equation}
For small values of $p_0$ and nonzero values of $q_0$, the
right-hand side of Eq.~(\ref{eq:cmplxrtplbndry}) tends to
diverge as $1/p_0^2$ just as would happen for $r_0 > 0,q_0=0$ because
the numerator does not vanish in the limit $p_0 \rightarrow 0$.
Furthermore, the values for the turning point marking the end of the
self-trapped region also have the same qualitative behaviour as in
the $r_0 >0, q_0=0$ case. Specifically, when $q_0 \neq 0,r_0 < 0$, the
values of the parameters at the turning point are:
$\Delta/V=3\sqrt{3}(1-\sqrt{1-q_0^2})/p_0^2,
\chi_c/V=\sqrt{2}\Delta/V$. Thus we see that the values of the
turning point diverge as $1/p_0^2$, similar to the case when
$r_0 >0, q_0 = 0$.
\par
It is possible to understand the qualitative similiarty between
the $r_0>0,q_0=0$ and $r_0<0,q_0\neq0$ cases. In doing so, we
can understand a general similarity between the results here
and those of Ref.~\cite{interplay}. The point is that
when $r_0<0,q_0 \neq 0$, in order to trap the system (either the DNLSE one or
its generalized cousin), one has to bring the system to a state that
is further away in character from the
stationary state of the system, marked by $r_0<0,q_0=0$, and thus closer
to one denoted by $r_0>0,q_0=0$. Thus both the cases
$r_0<0,q_0 \neq 0$ and $r_0>0,q_0=0$ are similar. What is
interesting is that this similarity is very deep and persists
right through into the strongly anharmonic regime.
\renewcommand{\theequation}{\arabic{section}.\arabic{equation}}
\setcounter{equation}{0}
\section{Logarithmic oscillator}
Along lines identical to those discussed above,
we have analyzed the free-selftrapped phase boundary for the system
of a quasiparticle in strong interaction with a logarithmic
oscillator discussed earlier in Ref.~\cite{occassnot}.
The counterpart of Eq.~(\ref{rotpol:gofp}) is given by
\beq
g(p) = \frac{8\chi_0^2}{\chi}e^{-\chi/2\chi_0}
\sinh^2\left(\frac{\chi p}{4 \chi_0}\right),
\label{sinh:gofp}
\eeq
where $\chi_0$ is a saturation parameter. The DNLSE
limit is recovered in the limit $\chi/\chi_0 \ll 1$. The
phase-boundary in the parameter space $\chi_0/V,\chi_c/V$ marking
the transition from free evolution to self-trapped behaviour is
given then by the expression
\beq
\frac{8\chi_0^2}{\chi_c}e^{-\chi_c/2\chi_0}
\sinh^2\left(\frac{\chi_c p}{4 \chi_0}\right) = 2 V (1+r_0)
\label{sinh:free-self}
\eeq
Since the phase-diagrams obtained for this system
are qualitatively similar to the ones obtained for the rotational polaron,
we do not discuss them in detail  here. The logarithmic oscillator,
besides possessing qualitatively features (similar to the
rotational polaron) like saturation of
nonlinearity also exhibits
 the same asymptotic forms for the phase boundaries and turning
points.
For example, the least saturation value required to support
self-trapping occurs for $r_0<0,p_0\rightarrow 0$ for
$\chi_0/V=e/(2\sqrt{2}),\chi_c=4 \chi_0$. The asymptotes
corresponding to $\chi_c/V,\chi_0/V\rightarrow\infty$,
are a bit different for $p_0 \ll 1$. In contrast to the
power-law type behaviour (with exponent 1.5) exhibited for the
rotational polaron case, the dependence of critical nonlinearity
$\chi_c$ on saturation parameter $\chi_0$ is a bit more complicated
and numerical evidence seems to indicate the exponent is only slightly
greater than 1. However, for $p_0=1$, we obtain a
result similar to that obtained for the rotational polaron, viz.,
that $\chi_c/V \sim (\chi_0/V)^2.$ For the case, $r_0>0$, the
available parameter space for self-trapping again shrinks like the
counterpart for the
rotational polaron. It is particularly simple to calculate
the asymptotic properties and turning point of the phase boundary (marking
the minimum saturation required for self-trapping) in the limit of small
values of $p_0$. In this regime, the turning point is marked by
$\chi_0/V = 8 e/p_0^2, \chi_c = \chi_0$. The lower asymptote (in
the DNLSE regime) is recovered: $\chi_c/V = 8/p_0^2, \chi_0/V \rightarrow
\infty$ whereas the upper asymptote in the anharmonic regime is
given by the implicit relation:
\beq
\chi_0 = \frac{4 \chi_c}{\log(\chi_c/V)p_0^2}
\; , \; \chi_c \rightarrow \infty.
\eeq
\renewcommand{\theequation}{\arabic{section}.\arabic{equation}}
\setcounter{equation}{0}
\section{Conclusion}
We have extended our analysis \cite{interplay} of the
effect of quantum-mechanical quasiparticle
phases on self-trapping by treating saturating dimers
 \cite{samos,kwh,wk,occassnot}.
We find that initial quasiparticle phases can compete or act with
nonlinearity as well as saturation effects. 
For the case $r_0 < 0$ (in contrast to $r_0 > 0$), the
region of available parameter space for
self-trapping increases with decreasing asymmetry of initial population,
$p_0$. In particular, as
reported in Ref.~\cite{wk},
there exists a minimum value of saturation
ratio $\Delta/V$ below which no self-trapping can occur for any value
of nonlinearity. What is interesting, however, is that this critical
value is the least when $r_0 < 0, p_0 \rightarrow 0$ and increases
for all other regimes. The diametrically opposite case of
$r_0 > 0, q_0 = 0,
p_0 \rightarrow 0$ yields a starkly different condition, viz., that
this critical value of saturation diverges as $1/p_0^2$.
A generalization to the case of complex initial
amplitudes gives results similar to the case when $r_0 > 0, q_0=0$.
We have also analyzed the logarithmic oscillator and obtained
qualitatively similar features as those shown by the rotational polaron.
We are thus led to conjecture that these features are perhaps universal to
dimers interacting with oscillator potentials that have saturating
nonlinearity.
\par
We gratefully acknowledge V. M. Kenkre for extremely useful discussions.

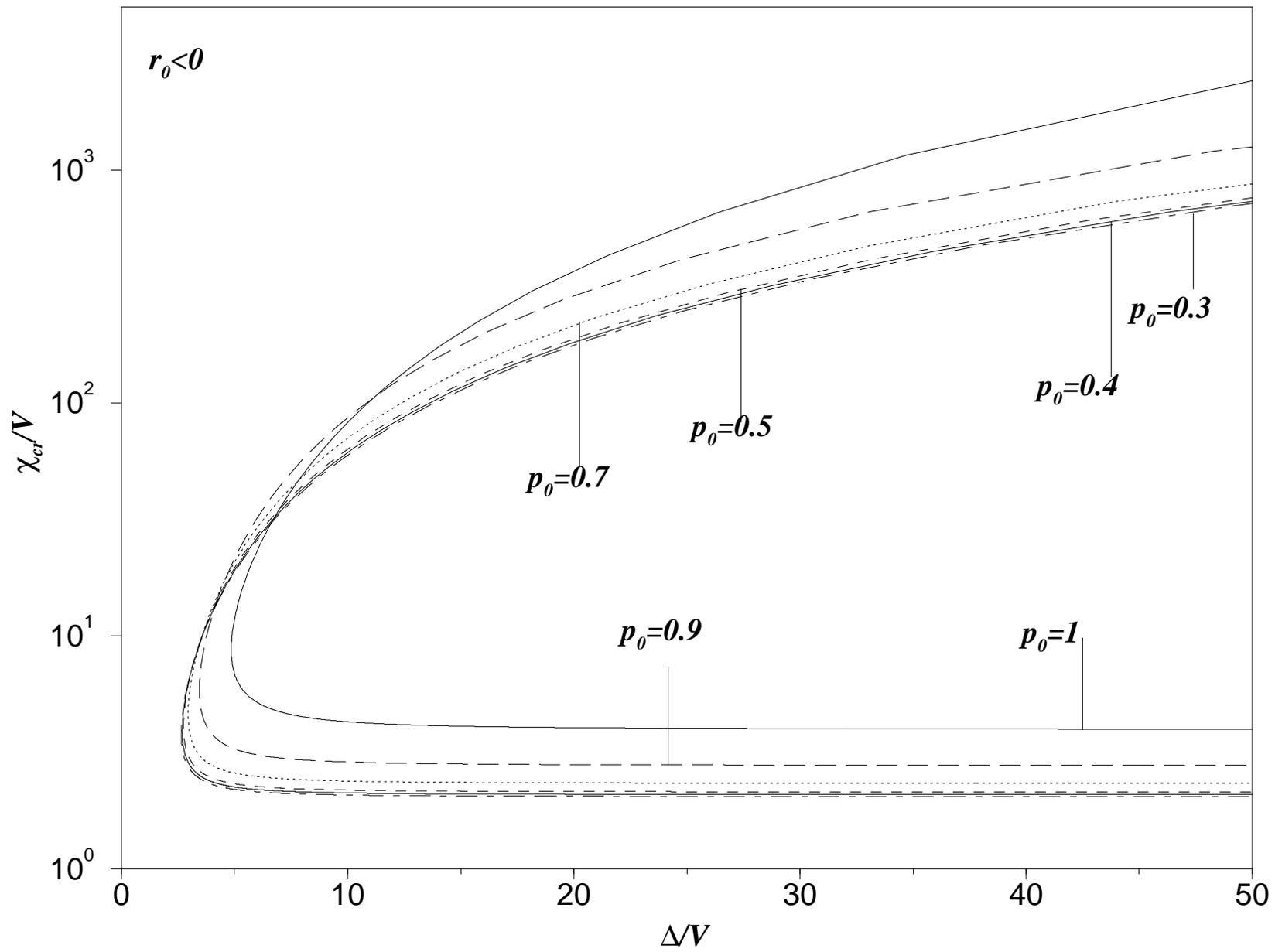
\begin{figure}
\caption{The critical value of nonlinearity ratio $\chi_c/V$, i.e., the
value required for self-trapping is plotted logarithmically against
the saturation ratio $\Delta/V$ for various values of
initial population difference $p_0$. Here, $r_0 <0,q_0=0$.}
\end{figure}
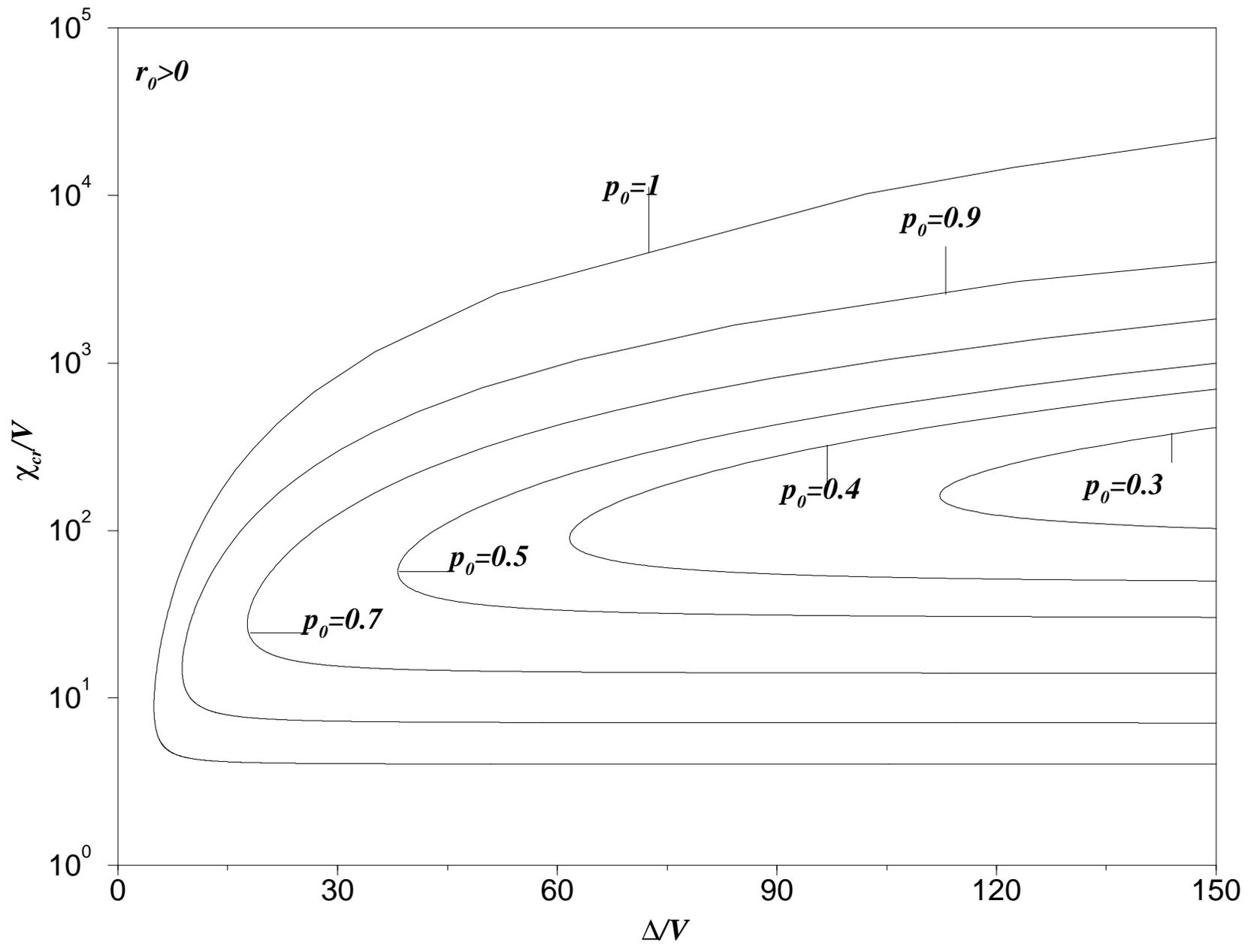
\begin{figure}
\caption{$\chi_c/V$ is plotted logarithmically against $\Delta/V$
for different values of $p_0$. In this figure, $r_0 >0,q_0=0$.}
\end{figure}
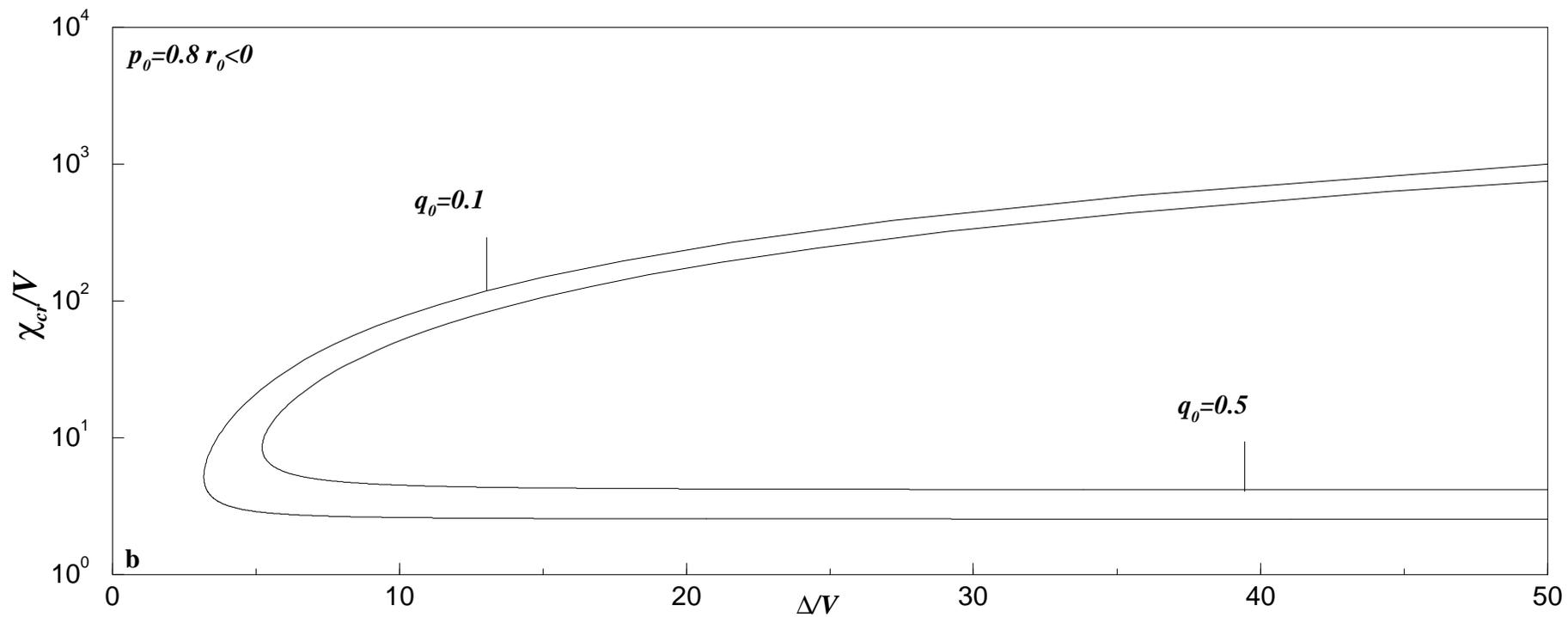
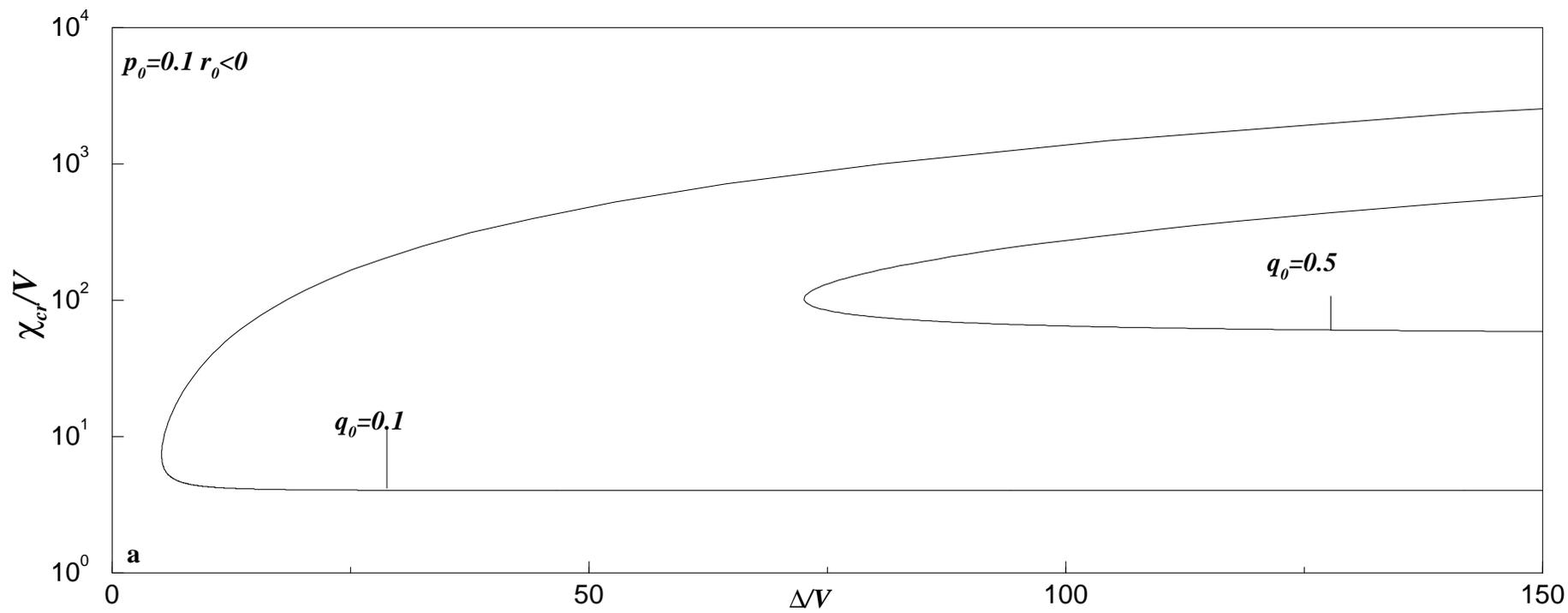
\begin{figure}
\caption{ $\chi_c/V$ is plotted logarithmically against $\Delta/V$ for
non-zero values of $q_0$. In (a), $p_0=0.1$ and in (b), $p_0=0.8$.
For both (a) and (b), $r_0<0$.}
\end{figure}


\begin{thebibliography}{10}

\bibitem{chandra}
S.~Chandrasekhar,
\newblock {\em Liquid Crystals},
\newblock Cambridge University Press, Cambridge, 1977.

\bibitem{samos}
V.~M. Kenkre,
\newblock The quantum nonlinear dimer and extensions,
\newblock in {\em Singular Behavior and Nonlinear Dynamics}, edited by
  S.~Pnevmatikos, T.~Bountis, and S.~Pnevmatikos, World Scientific, Singapore,
  1989.

\bibitem{kwh}
V.~M. Kenkre, H.-L. Wu, and I.~Howard,
\newblock Phys. Rev. B {\bf 51}, 15841 (1995).

\bibitem{wk}
H.-L. Wu and V.~M. Kenkre,
\newblock Phys. Lett. A {\bf 199}, 61 (1995),
\newblock see also H.-L. Wu, Ph. D. thesis, (University of New Mexico, 1990),
  p. 126-148, unpublished.

\bibitem{occassnot}
V.~M. Kenkre, M.~F. J\mbox{\o}rgensen, and P.~L. Christiansen,
\newblock Physica D {\bf 90}, 280 (1996).

\bibitem{interplay}
S.~{R}aghavan, {V}.~{M}.~{K}enkre, and {A}.~{R}.~{B}ishop, to appear in {P}hys.
  {L}ett. {A}.

\bibitem{vit}
D.~Vitali, P.~Allegrini, and P.~Grigolini,
\newblock Chem. Phys. {\bf 180}, 297 (1994).

\bibitem{sbkr}
M.~I. Salkola, A.~R. Bishop, V.~M. Kenkre, and S.~Raghavan,
\newblock Phys. Rev. B {\bf 52}, R3824 (1995).

\bibitem{krbs}
V.~M. Kenkre, S.~Raghavan, A.~R. Bishop, and M.~I. Salkola,
\newblock Phys. Rev. B {\bf 53}, 5407 (1996).

\bibitem{kc}
V.~M. Kenkre and D.~K. Campbell,
\newblock Phys. Rev. B {\bf 34}, 4959 (1986).

\bibitem{scatter}
V.~M. Kenkre and G.~P. Tsironis,
\newblock Phys. Rev. B {\bf 35}, 1473 (1987).

\bibitem{ktc}
V.~M. Kenkre, G.~P. Tsironis, and D.~K. Campbell,
\newblock in {\em Nonlinearity in Condensed Matter}, edited by A.~R. Bishop,
  D.~K. Campbell, P.~Kumar, and S.~Trullinger, Springer, Berlin, 1989.

\bibitem{tk}
G.~P. Tsironis and V.~M. Kenkre,
\newblock Phys. Lett. A {\bf 127}, 209 (1988).

\bibitem{kt}
V.~M. Kenkre and G.~P. Tsironis,
\newblock Chem. Phys. {\bf 128}, 219 (1988).

\end{thebibliography}
\end{document}